\documentstyle[12pt]{article}

\def\dspace{\baselineskip = .30in}

\def\VEV#1{\left\langle #1\right\rangle}
\def\Tilde#1{\widetilde {#1}}

\begin{document}
\title{A Solution to the Small Phase Problem of
Supersymmetry\thanks{Supported in part by Department of Energy Grant
\#DE-FG02-91ER406267}}

\author{{\bf K.S. Babu} and {\bf S.M. Barr}\\ Bartol Research Institute\\
University of Delaware\\ Newark, DE 19716}

\date {bA-93-48}
\maketitle
\begin{abstract}

\dspace

It is a well--known problem that in supersymmetric models there are new
CP--violating phases which, if unsuppressed, would give a neutron
electric dipole moment $10^2$ to $10^3$ times the present
experimental limit.  Here we propose that these new phases
are suppressed by CP invariance, which is broken spontaneously
at a high scale and that this breaking shows up at low energies
only through a universal phase of the gaugino masses.  It is shown
that this can well fit both $\epsilon$ and $\epsilon^\prime$ of the
neutral Kaon system.  The electric dipole moments of the neutron
and the electron should be not much below present limits.
A model incorporating these ideas in a very economical way is presented.

\end{abstract}
\newpage
\dspace

In this letter we propose a model of CP violation that solves the
small phase problem of supersymmetry and has testable
low--energy consequences.  The idea is that CP violation arises
spontaneously at a high scale and is communicated to the low--energy
world through a common phase of the gaugino--masses.  All
low--energy CP violation would be a consequence of this one
non--vanishing phase angle.  The $\epsilon$ parameter of the Kaon system
would arise primarily from the phase of the gluino mass through the box
diagram [1] shown in Fig. 1.  As we
shall see, to fit $\epsilon$ the phase of the gaugino mass must be
$\stackrel{_>}{_\sim} 3 \times 10^{-3}$,
and the gluino should be relatively light
($M_{\Tilde{g}} \stackrel{_<}{_\sim} 500~GeV$).
This leads to
a value of $|\epsilon^\prime/\epsilon| \simeq (1~{\rm to}~3) \times 10^{-3}$
(modulo hadronic matrix element uncertainties) which
arises in the model dominantly via the gluino penguin
graph of Fig. 2.  Electric dipole moments (edm) [2] of the neutron
$(d_n$) [3,4] and
electron ($d_e$) [4]
would be induced by the one--loop diagrams shown in Figs. 3 and
4, which turn out to be not far
below the present experimental limits.

The problem that is solved by this idea is the tendency of the neutron
edm arising from Fig. 3 to come out about a factor of $10^2$ to
$10^3$ too large in models with low--energy supersymmetry [3,4].  In SUSY
models there are new sources of CP violation in the $A$ and $B$
parameters, the $\mu$ parameter, and the gaugino masses.  If CP is
explicitly broken there is no reason, in general, why these phases
should be small.  If one assumes, as is natural, that these phases are
of order unity and that the various as--yet--unobserved superparticles
(gluino, squarks) have masses around $100~GeV$, then one finds that
$d_n \sim 10^{-22}$ e-cm, to be compared with present upper limit of
$d_n \le 10^{-25}$e-cm.

One solution to this well--known difficulty is to assume that CP is a
spontaneously broken symmetry.  Then CP--violating parameters are
finite, calculable, and, if they arise radiatively, naturally small.
This general approach to the problem, which is not new, raises two
issues.  The first is that spontaneous CP violation leads to cosmic
domain walls.  These can be rendered harmless if they are ``inflated
away''.  This requires that CP be broken at scales larger than the
reheating temperature, which argues for the scale of spontaneous CP
violation to be much higher than $M_W$.

The second issue is how the CP
violation arising spontaneously at large scales is ``fed down'' to the
Kaon system.  Several ``feeding--down'' mechanisms have been proposed in the
literature [5,6,7].  Those suggested in Ref. 5 and 6 were motivated by the
desire to solve the $\theta$--problem (the strong CP problem) using
spontaneous CP violation and were therefore necessarily somewhat
intricate.  In any event, it was shown in Ref. 8 that these non--axion
approaches to the $\theta$--problem are fraught with difficulties in the
context of supersymmetry.
In our model the $\theta$--problem is solved by the Peccei--Quinn
mechanism [9], in particular by the KSVZ invisible axion [10], and thus our
feeding--down mechanism can be much more straightforward than the
proposals in Ref. 5
and 6.

The essential idea is that CP is spontaneously broken by the
vacuum--expectation values (VEVs) of certain gauge--singlet scalar
fields, which we will call $S_i$, $\VEV{S_i} \gg M_W$.  These VEVs
give large complex masses to some vector--like fermions (needed anyway to
realize the KSVZ invisible axion) which
are non--singlet under the gauge group.  These fermions,
which we will denote $Q+Q^c$, do not mix with the known quarks and
leptons (owing to their PQ charges).
When $Q$ and $Q^c$ are integrated out, the masses of the
gauginos that couple to them
will acquire a CP--violating phase at one--loop which is naturally of
order $10^{-2}~{\rm to}~10^{-3}$.

If grand unification is assumed and it is also assumed that the mass
of $Q$ and $Q^c$ comes predominantly from these gauge--singlet
contributions of $\VEV{S_i}$, then the phases of the masses of the
$SU(3)_C,~SU(2)_L$ and $U(1)_Y$ gauginos will be very nearly equal.  In
this scenario the only significant CP violating phase in the low energy
theory is this common gaugino mass--phase, and thus all low--energy
CP--violation phenomenology is controlled by one parameter.

A model will now be presented which shows how this idea can be
implemented in a particularly economical way in which the same sector
does the breaking of both the CP invariance and the Peccei--Quinn
symmetry.

Consider a SUSY $SU(5)$ model in which, in addition to the known matter
fields, there is a \{5\} + \{$\bar{5}$\} with Peccei--Quinn charge $-1/2$.
Denote these $Q+Q^c$.  Coupling to these are two $SU(5)$--singlet
superfields $S_1$ and $S_2$, both with Peccei--Quinn charge of $+1$.
A third singlet, $S_3$, carries $PQ$ charge of $-2$.  The
superpotential of this sector is given by
\begin{equation}
W = Q^c Q \left(f_1 S_1+f_2S_2\right) + \left(a_{11}S_1^2+a_{12}S_1S_2 +
a_{22} S_2^2\right)S_3
\end{equation}
where by CP invariance all the parameters are real.  Consider the
case where $\VEV{Q} = \VEV{Q^c} = \VEV{S_3} = 0$.  Then
$F_{S_1}= F_{S_2} = F_Q = F_{\overline{Q}}=0$ are automatically satisfied.
The breaking of CP can arise as a result of the $F_{S_3}=0$ equation,
\begin{equation}
a_{11}S_1^2+a_{12}S_1S_2 + a_{22} S_2^2 = 0~,
\end{equation}
which is solved for
\begin{equation}
\VEV{S_2} = k\VEV{S_1},~~~ k \equiv \left(-a_{12} \pm \sqrt{a_{12}^2-4a_{11}
a_{22}}\right)/(2a_{22})~.
\end{equation}
If $a_{12}^2 - 4 a_{11}a_{22} < 0$, then there will be a non--trivial
relative phase between $\VEV{S_1}$ and $\VEV{S_2}$, which breaks CP
invariance.

Note that Eq. (3) leaves the magnitude of $\VEV{S_1},~\VEV{S_2}$
undetermined.  It
is easy to fix them at the desired $ PQ$ scale in several ways.  For
example, an extra term in the superpotential, $\left(S_1\overline{S_1}
-M_{PQ}^2\right)X$, where $\overline{S_1}$ and $X$ are singlet fields with PQ
charges of $-1$ and $0$ respectively, would lead to
$\VEV{S_1}\VEV{\overline{S}_1} = M_{PQ}^2$ from the $F_X=0$ equation.  The
soft SUSY term $m^2(|S_1|^2+|\overline{S_1}|^2)$ will minimize the
potential for $\VEV{S_1} = \VEV{\overline{S_1}} = M_{PQ}$.
For $a_{11},~a_{22}$ and $a_{12}$ all of the same order,
one will have $\VEV{S_2}$ being also of order $M_{PQ}$.  Hence the same
fields that break $U(1)_{PQ}$, namely $S_1$ and $S_2$, break CP
spontaneously.  The relative phase, $[arg\VEV{S_1}-arg\VEV{S_2}]$, is
the source of all CP violation in the model, while the phase
$[arg\VEV{S_1}+arg\VEV{S_2}]$ is essentially the invisible axion.

The ``quarks'' $Q+Q^c$ that implement the KSVZ
axion idea are the means of feeding CP violation to the observable
low--energy world.
The feeding down occurs through the (s)quark--loop contribution to the
gaugino mass shown in Fig. 5.  There will be a soft--SUSY breaking term
for the squarks of the form
\begin{equation}
V_{\rm Soft} = A_1 f_1 (Q^c Q S_1) + A_2 f_2 (Q^cQS_2) + H.c.
\end{equation}
Thus the phase appearing in the squark mass insertion in Fig. 5 is
\newline $arg\left(A_1f_1\VEV{S_1}^*+A_2f_2\VEV{S_2}^*\right)$
while that appearing in the quark mass insertion is
$arg\left(f_1\VEV{S_1}+f_2\VEV{S_2}\right)$.
If $A_1$ and $A_2$ were equal these phases would cancel and the
one--loop contribution to the gaugino mass would be real.  However,
$A_1 \ne A_2$ in general.  Even if $A_1 = A_2$ at the Planck scale (as
is expected in supergravity models), they run differently if $a_{11}
\ne a_{22}$ and would be significantly different at the Peccei--Quinn
scale, $M_{PQ}$, which we assume to be between $10^{10}$ and $10^{12}$
GeV.  One finds for the phase of the gaugino mass
\begin{equation}
arg(M_{\Tilde{g}}) \equiv \phi =
{{\alpha_G} \over {8 \pi}} \left({{A_1-A_2} \over
{M_{1/2}}}\right)\left\{{{2 f_1f_2|k|{\rm sin}\Delta}\over {f_1^2+
f_2^2|k|^2+2f_1f_2|k|{\rm cos}\Delta}}\right\}~.
\end{equation}
Here $M_{1/2}$ is the common gaugino mass at $M_{GUT}$, $\alpha_G$ the
gauge coupling strength at $M_{GUT}$ and $\Delta$ the phase of $k$ in
Eq. (3).

A number of remarks are now in order:

(i) If one neglected the effects of the running of the
parameters between $M_{GUT}$ and $M_{PQ}$, then the one--loop
calculation of $arg(M_{\Tilde{g}})$ given in eq. (5) would be manifestly
$SU(5)$ invariant, and the phases of the gluino, the Wino and the Bino would be
all the same.  Interestingly, and slightly non--trivially, this result
remains true to one--loop order in the RGE even when the running is
taken into account, as explained below.

In the momentum range $M_{PQ} \le \mu \le M_{GUT}$, since $SU(5)$
symmetry is not exact,
the first term in eq. (1) will split into two pieces, a color--triplet
($\Omega$) part and an $SU(2)$--doublet ($L$) part:
\begin{equation}
W = \Omega^c \Omega (f_1S_1+f_2S_2) + {L}^c L
(f_1^\prime S_1 + f_2^\prime S_2) + ....
\end{equation}
Similarly, the soft SUSY breaking terms of eq. (4) will split into
\begin{eqnarray}
V_{\rm Soft} = A_1 f_1 (\Omega^c \Omega S_1) + A_2 f_2 \Omega^c \Omega
S_2 + A_1^\prime f_1^\prime (L^c L S_1) + A_2^\prime f_2^\prime (L^c L S_2) +
H.c.
\end{eqnarray}
At and above $M_{GUT}$, one has $f_1=f_1^\prime, ~f_2=f_2^\prime$, and
$A_1=A_1^\prime,~A_2=A_2^\prime$.  From the renormalization group
equations for the various parameters of the
model we find that in the momentum range between $M_{GUT}$ and $M_{PQ}$,
\begin{equation}
{d \over {dt}} (A_1-A_2) = {d \over {dt}}(A_1^\prime-A_2^\prime);~~
{d \over {dt}} \left({{f_1} \over {f_2}}-{{f_1^\prime} \over
{f_2^\prime}}\right) \propto
\left({{f_1} \over {f_2}}-{{f_1^\prime} \over
{f_2^\prime}}\right)~.
\end{equation}
It follows from the above that $(A_1-A_2) = (A_1^\prime-A_2^\prime)$ and
$f_1/f_2 = f_1^\prime/f_2^\prime$ at all scales.  Combining with the
scaling of gaugino masses, namely, $(\alpha_i/M_i) = (\alpha_G/M_{1/2})$, we
arrive at the result that the phase of all the gauginos are identical at
the $PQ$ scale even
after $SU(5)$ symmetry breaking.  They will then remain to
be equal down to the weak scale.

(ii) The gaugino phase, $\phi$, can easily be $\stackrel{_>}{_\sim} 3 \times
10^{-3}$ which is
what is typically required (as will be seen below) to generate
$\epsilon$ in the K meson system
from the graph of Fig. 1.  However, Eq. (5) shows that
$\phi$ large enough to fit $\epsilon$ requires that the gaugino masses
not be too large, a point that is important for the expected
magnitudes of the neutron and electron edms.  Taking $\alpha_G \simeq
1/28$ and noting that the magnitude of the function
in the curly bracket of Eq. (5) is less than unity, we see that a phase
angle $\phi \stackrel{_>}{_\sim} 3 \times 10^{-3}$ requires $(A_1-A_2)/M_{1/2}
\stackrel{_>}{_\sim} 2.1$.  This is both
an upper limit on the gaugino mass and a lower limit on the $A$
parameter.  Solving the RGE for the $A$ parameters we found that
$(A_1-A_2) \stackrel{_<}{_\sim} 0.7 A_0$, where $A_0$
is the universal $A$ parameter at the Planck scale.  For $A_0= 500~GeV$,
we see that $M_{1/2} \stackrel{_<}{_\sim} 170 ~GeV$, which after RGE
corrections correspond to $M_{\Tilde{g}} \stackrel{_<}{_\sim} 500~GeV$
at the weak
scale.  The experimental lower limit on $M_{\Tilde{g}}
\stackrel{_>}{_\sim} 150 ~GeV$
implies that $A_0 \stackrel{_>}{_\sim} 150~GeV$,
which could have important consequences
for the electroweak symmetry breaking.

(iii) In general complex VEVs can induce phases in the $A$ and
$B$ parameters of the ordinary sector at tree--level as noted in Ref. 6.
This can be avoided if $W(\phi_i)|_{\phi_i=\VEV{\phi_i}}$ is
real.  Since $\VEV{Q^c} = \VEV{Q} = \VEV{S_3} = \VEV{X} = 0$, this
condition is trivially satisfied for Eq. (1) since $W$ evaluated at
$\phi_i = \VEV{\phi_i}$ vanishes identically.

(iv) There is no reason to expect any other phase than the
common gaugino phase to be significantly large at low energy.  For
example, in the KSVZ axion model such as this, where the known quarks
and leptons and the Higgs superfields $H_1$ and $H_2$ have vanishing
Peccei--Quinn charge, the $Q^c-Q-S_i$ sector is quite separate from the
sector of ordinary matter (except through their coupling to the
gauge/gaugino particles).  Thus no one--loop diagram involving
$\VEV{S_i}$ contributes to $\mu$ or $B\mu$ or to the Yukawa couplings of
the known quarks and leptons.

We now turn to the evaluation of the CP violating parameters $\epsilon$,
$\epsilon^\prime/\epsilon$ and the neutron and
the electron electric dipole moments in the model.  The $\Delta S = 2$
CP violating effective Hamiltonian is obtained from the gluino box graph
of Fig. 1 (the $SU(2)$ gaugino box graph is suppressed by a factor of
$\sim 30$ relative to the gluino box graph and thus is negligible).
\begin{eqnarray}
{\cal H}_{\rm eff}^{\Delta S = 2} = {{ \alpha_s^2}\over {10 M_{sq}^2}}
\delta_{LR}^2 {\rm sin}2\phi~ x f(x) \left[{7 \over 3}
\overline{s}_{R \alpha} d_{L}^\alpha \overline{s}_{R\beta} d_L^\beta
+
{5 \over 9}\overline{s}_{R \alpha}d_L^\beta
\overline{s}_{R \beta} d_L^\alpha\right] - (L \leftrightarrow R).
\end{eqnarray}
Here $M_{sq}$ is the (common) squark mass, $\alpha, \beta$ are the
color indices, $\phi$ is the phase of the gluino mass [Eq. (5)],
$x=M_{\Tilde{g}}^2/M_{sq}^2$
and the function $f(x)$ is defined as
\begin{equation}
f(x) = {10 \over {3(1-x)^5}}\left(9 x + 9 x^2-x^3-6 {\rm ln}x -18x
{\rm ln}x-17\right)
\end{equation}
with $f(1)=1$.  The parameter $\delta_{LR}$ is defined to
be
$\delta_{LR} = m^2_{\Tilde{d}_L \Tilde{s}_R}/M_{sq}^2$.  Since the
mass--splitting among squarks is constrained phenomenologically to be
small, we have treated the $\Tilde{d}_L-\Tilde{s}_R$ mass insertion
(denoted by $m^2_{\Tilde{d}_L \Tilde{s}_R}$) in
Fig. 1 as small perturbation.  Note that other gluino graphs which
do not involve $\Tilde{d}_L\Tilde{s}_R$ mass insertions (e.g.,
one with $\Tilde{d}_L\Tilde{s}_L$ mass insertion) are real and do
not contribute to $\epsilon$.  This simplification is a consequence of
the fact that only the gluino mass has a non--vanishing phase.  The
contribution to Re$M_{12}$ (or $\Delta m_K$)
from Fig. 1 is obtained by the interchange
sin$2\phi \leftrightarrow {\rm cos}2\phi$ and taking a relative plus sign
between the $(LR)$ and $(RL)$ terms.

The $\epsilon$ parameter evaluated from Eq. (9) is given by
\begin{equation}
|\epsilon| = {5 \over {54 \sqrt{2}}}
B \eta {{\alpha_s^2} \over {M_{sq}^2}}
(\delta_{LR}^2-\delta_{RL}^2){\rm sin}2\phi~ x f(x)
{{f_K^2m_K}
\over {\Delta m_K}}\left({{m_K}\over {m_d+m_s}}\right)^2~,
\end{equation}
where in the vacuum saturation method of evaluating the $K-\overline{K}$
matrix element $B$ would be 1 by definition.
$\eta$ is the QCD correction factor from $M_{sq}$ to
the hadronic scale.  If $\alpha_s$ in Eq. (11) is
evaluated at the $\mu=M_{sq}$, then $\eta \simeq 1.8 [11]$ for
$\alpha_s \simeq 0.12$.
$f_K \simeq 165 ~MeV$ is the Kaon decay constant.
The function $xf(x)$ is slowly varying with its value
changing from $1$ for $x=1$ to $1.1$ for $x=0.1$.
Using $m_s=150~MeV,~m_d=10~MeV$ and $x=1$, we obtain by fitting $|\epsilon| =
2.3 \times 10^{-3}$,
\begin{equation}
B {\rm sin}2\phi (\delta_{LR}^2-\delta_{RL}^2) \simeq 3.2 \times
10^{-9} \left({ {M_{sq}} \over {300~GeV}}\right)^2~.
\end{equation}
In our spontaneous--CP violation mechanism, the phase angle $\phi$ is
naturally of order $3 \times 10^{-3}$, so that the larger of the
mass--splittings,
$\delta_{LR}$ or $\delta_{RL}$ must be $\sim 10^{-3}$, assuming that
they are not accidentally close in value, (i.e., assuming that one of them
dominates).
Demanding that the contribution from the real part of Fig. 1 not be larger than
the experimental value of $\Delta m_K$, we obtain
\begin{equation}
B {\rm cos}2\phi (\delta_{LR}^2+\delta_{RL}^2) \stackrel{_<}{_\sim}
1.0 \times 10^{-6} \left({{M_{sq}} \over {300~GeV}}\right)^2~.
\end{equation}
{}From Eq. (12) and (13), we obtain the constraint $\phi
\stackrel{_>}{_\sim} 3 \times 10^{-3}$.

The dominant contribution to the
$\Delta S = 1$ CP violating effective Hamiltonian arises from the gluino
penguin graph of Fig. 2.  (The $U(1)_Y$ gaugino penguin contribution is
two orders of magnitude smaller, the $SU(2)$ gauginos do not
contribute directly.  There are graphs involving $\Tilde{W}^+
\Tilde{H}_1^-$ mixing, but these are an order of magnitude smaller.)
Evaluating Fig. 2 we obtain
\begin{equation}
{\cal H}_{\rm eff}^{\Delta S = 1} = \left({7 \over 9}\right)
{1 \over {256 \pi^2}}
{{g_3^3}\over
{M_{sq}}}(\delta_{LR}-\delta_{RL}){\rm sin}\phi \sqrt{x} g(x)
\left[\overline{s}\lambda^a i \sigma_{\mu \nu} (1-\gamma_5)d
G_{\mu \nu}^a + H.c. \right]
\end{equation}
where
\begin{equation}
g(x) = {{2\left(2+3x-6x^2+x^3+6x{\rm ln}x\right)}\over{(1-x)^4}}
\end{equation}
with $g(1) = 1$.  We use the bag model calculation of Ref. (12) to
evaluate the hadronic matrix element in Eq. (15) and obtain
\begin{equation}
{{\epsilon^\prime} \over {\epsilon}} =
5.6 \times 10^2 B^\prime\eta^\prime{\rm sin}\phi
\sqrt{x}g(x)(\delta_{LR} - \delta_{RL}) \left({{300~GeV} \over {M_{sq}}}
\right)~.
\end{equation}
Here $\eta^\prime$ is the QCD correction factor, $\eta^\prime = [\alpha_s(\mu)/
\alpha_s(M)]^{0.92} \simeq 3$, where $\mu \sim 1~GeV$ and
$\alpha_s(\mu) \simeq 0.4$ has been used.
$B^\prime$ is a factor introduced to parameterize the uncertainty in
the matrix element and is defined to be 1 if the Bag model matrix
elements given in Ref. (12) are exact.
Combining Eq. (12) with Eq. (16), we obtain the prediction (for $x=1$)
\begin{equation}
{{\epsilon^\prime} \over {\epsilon}} \simeq 2.7 \times 10^{-6} \left(
{{B^\prime}\over {B}}\right)\left({{M_{sq}}\over {300~GeV}}\right)
{1 \over {\delta_{LR}+\delta_{RL}}}~.
\end{equation}
If we make the reasonable assumption that either $\delta_{LR}$ or
$\delta_{RL}$ dominates the squark mass-splitting, we obtain
$|\epsilon^\prime/\epsilon| \simeq (1 ~{\rm to}~3)\times 10^{-3}$.  This
is clearly in the range suggested by experiments.  Note that the sign of
$\epsilon^\prime/\epsilon$ is not predicted in our model.

In supergravity models, if the minimal supersymmetric
spectrum extends all the way upto the Planck
scale, the squark mass--splitting will be too small for Eq. (11) to
account for $\epsilon$.
However, the MSSM spectrum is not expected to hold all the
way to $M_{\rm Pl}$, since the GUT threshold will in general bring in new
effects [13].  A simple example is the realization of the
see--saw mechanism for neutrino masses.  Between the GUT scale and
the Planck scale, the
Dirac and Majorana neutrino matrices,
with their elements not necessarily small, will
contribute to the evolution of the squark mass matrix.  The running in
this short momentum range can result
in relatively large values of the mass--splitting.  A typical diagram
which can generate $\Tilde{d}_L\Tilde{s}_R$ mixing via the neutrino
Dirac mass matrix in $SU(5)$ is shown in Fig. 6, which can lead to $\delta_{LR}
\sim (10^{-4}~{\rm to}~10^{-3})$.  It has also been emphasized [8] that the
squark mass--degeneracy in supergravity models, in the absence of
additional symmetries, will naturally be
$\delta m_{sq}^2/M_{sq}^2 \sim {\cal O}(\alpha)$.  In realistic
string compactification scenarios, the squark degeneracy is indeed
of this order.

One of the most interesting consequences of our fundamental hypothesis
that all low--energy CP violation is the result of a common
gaugino--phase is that both $d_n$ and $d_e$ are to be expected at a
measurable level.  Of course there are large hadronic uncertainties in
$d_n$, but it is generally estimated that, with phases of order unity
and sparticle masses of order 100 GeV, $d_n$ from Fig. 3 will be about
$10^2$ to $10^3$ times the experimental bound, as noted earlier.  Since
we require our gluino phase to be $\stackrel{_>}{_\sim}3 \times 10^{-3}$
to fit $\epsilon$, it
is natural to expect $d_n$ to lie not far below the present bound.

Again, there are too many presently unknown SUSY parameters involved to
allow a calculation of the electron edm.  However, as noted in Ref. 14,
if all the superparticles have comparable masses and the gaugino phases
are all comparable, one would expect that $d_e \sim 10^{-2} d_n$.  Thus
$d_e$ should lie not far below $10^{-27}$e-cm.

\section*{References}

\begin{enumerate}
\item J.F. Donoghue, H.P. Nilles and D. Wyler, Phys. Lett. {\bf 128B},
55 (1983); M.J. Duncan, Nucl. Phys. {\bf B221}, 285 (1983); F. Gabbiani
and A. Masiero, $ibid$., {\bf B322}, 235 (1989); J.S. Hagelin, S. Kelley
and T. Tanaka, Preprint MIU-THP-92/60 (1993).
\item For reviews, see eg., S.M. Barr and W.A. Marciano, in {\it CP
Violation}, (ed. C. Jarlskog, World Scientific, 1989); W. Bernreuther
and M. Suzuki, Rev. Mod. Phys. {\bf 63}, 313 (1991).
\item W. Buchmuller and D. Wyler, Phys. Lett. {\bf 121B}, 321 (1982);
J. Polchinski and M. Wise, $ibid.$, {\bf 125B}, 393 (1983);
E. Franco and M. Mangano, $ibid$., {\bf 135B}, 445 (1984).
\item F. del Aguila, M.B. Gavela, J.A. Grifols and A. Mendez, Phys.
Lett. {\bf 126B}, 71 (1983).
\item A. Dannenberg, L. Hall and L. Randall, Nucl. Phys. {\bf B271}, 574
(1986).
\item S.M. Barr and A. Masiero, Phys. Rev. {\bf D38},, 366 (1988);
S.M. Barr and G. Segre, $ibid$. {\bf D48}, 302 (1993).
\item A. Pomarol, Phys. Rev. {\bf D47}, 273 (1993); R. Garisto and G.
Kane, Preprint TRI-PP-93-1 (1993).
\item M. Dine, R. Leigh and A. Kagan, Preprint SLAC-PUB-6090 (1993).
\item R.D. Peccei and H.R. Quinn, Phys. Rev. Lett. {\bf 38}, 1440
(1977).
\item J.E. Kim, Phys. Rev. Lett. {\bf 43}, 103 (1979); M.A. Shifman,
A.I. Vainshtein and V.I. Zakharov, Nucl. Phys. {\bf B166}, 199 (1981).
\item G. Ecker and W. Grimus, Z. Phys. {\bf C30}, 293 (1986).
\item J.F. Donoghue, E. Golowich, B.R. Holstein and W.A. Ponce, Phys.
Rev. {\bf D23}, 1213 (1981).
\item L.J. Hall, V.A. Kostelecky and S. Raby, Nucl. Phys. {\bf B267},
415 (1986); F. Gabbiani and A. Masiero, Ref. 1.
\item W. Bernreuther and M. Suzuki, Ref. 2.
\end{enumerate}

\section*{Figure Captions}
\begin{itemize}

\item Fig. 1.  A box diagram whereby the phase of the gluino mass contributes
to $Im(M_{12})$ in the neutral Kaon system.

\item Fig. 2.  The gluino penguin graph contribution to
$|\epsilon^\prime/\epsilon|$.

\item Fig. 3.  A contribution to the edm of the $d$--quark coming from the
phase of the gluino mass.  A similar diagram exists for the $u$--quark.
These in turn induce an edm of the neutron of comparable magnitude.

\item Fig. 4.  A contribution to the edm of the electron arising from the
phase of the photino mass.  There are several other diagrams involving
neutralinos and charginos.

\item Fig. 5.  The diagram by which the gaugino masses acquire a phase of
order $3 \times 10^{-3}$.

\item Fig. 6.  A diagram contributing to the $\Tilde{d}_L\Tilde{s}_R$
squark mass insertion proportional to the neutrino Dirac mass matrix.
\end{itemize}

\end{document}